\documentclass{WileyMSP-template}

\usepackage{comment}
\usepackage{cite}
\usepackage{colortbl}
\usepackage{makecell}
\usepackage{xcolor}
\usepackage{tabu}
\newcounter{magicrownumbers}
\newcommand\rownumber{\stepcounter{magicrownumbers}\arabic{magicrownumbers}}
\usepackage[onehalfspacing]{setspace}

\begin{document}


\title{More than just light management -- The multiple advantages of nano- and micro-textures in perovskite solar cells}

\maketitle


\author{Guillermo Mart\'inez-Denegri}
\author{Christiane Becker*}



\begin{affiliations}
Dr. G. Mart\'inez-Denegri, Prof. Dr. C. Becker\\
Division Solar Energy, Helmholtz-Zentrum Berlin f\"ur Materialien und Energie GmbH, Kekul\'estr. 5, 12489 Berlin, Germany\\
Email Address: christiane.becker@helmholtz-berlin.de


\end{affiliations}



\begin{abstract}

Perovskite-based solar cells have undergone rapid improvements over the last decade enabling highest power conversion efficiencies of single-junction and multi-junction devices. The implementation of nano- or micro-textures has played a major role in this development due to their ability to minimize reflection losses, maximize light trapping, and hence increase light harvesting in the active layer. In the slipstream of these advances, it has become apparent that nano- and micro-textures can also have many other collateral benefits beyond light management. In this article, very recently reported texture-related benefits for perovskite-based solar cells different from light management are reviewed. These are namely (1) improved film wetting from perovskite-solution, (2) enhanced perovskite crystallinity in terms of grain size, crystal orientation and phase homogeneity, (3) enhanced carrier extraction leading to higher open-circuit voltages, and potentially (4) increased mechanical stability upon bending in flexible devices and reduced residual stress. This survey helps to understand textures as a holistic concept to further improve perovskite-based optoelectronic devices such as solar cells, photodetectors and light emitting diodes.

\end{abstract}


\section{Introduction}

\begin{figure}
  \includegraphics[width=\linewidth]{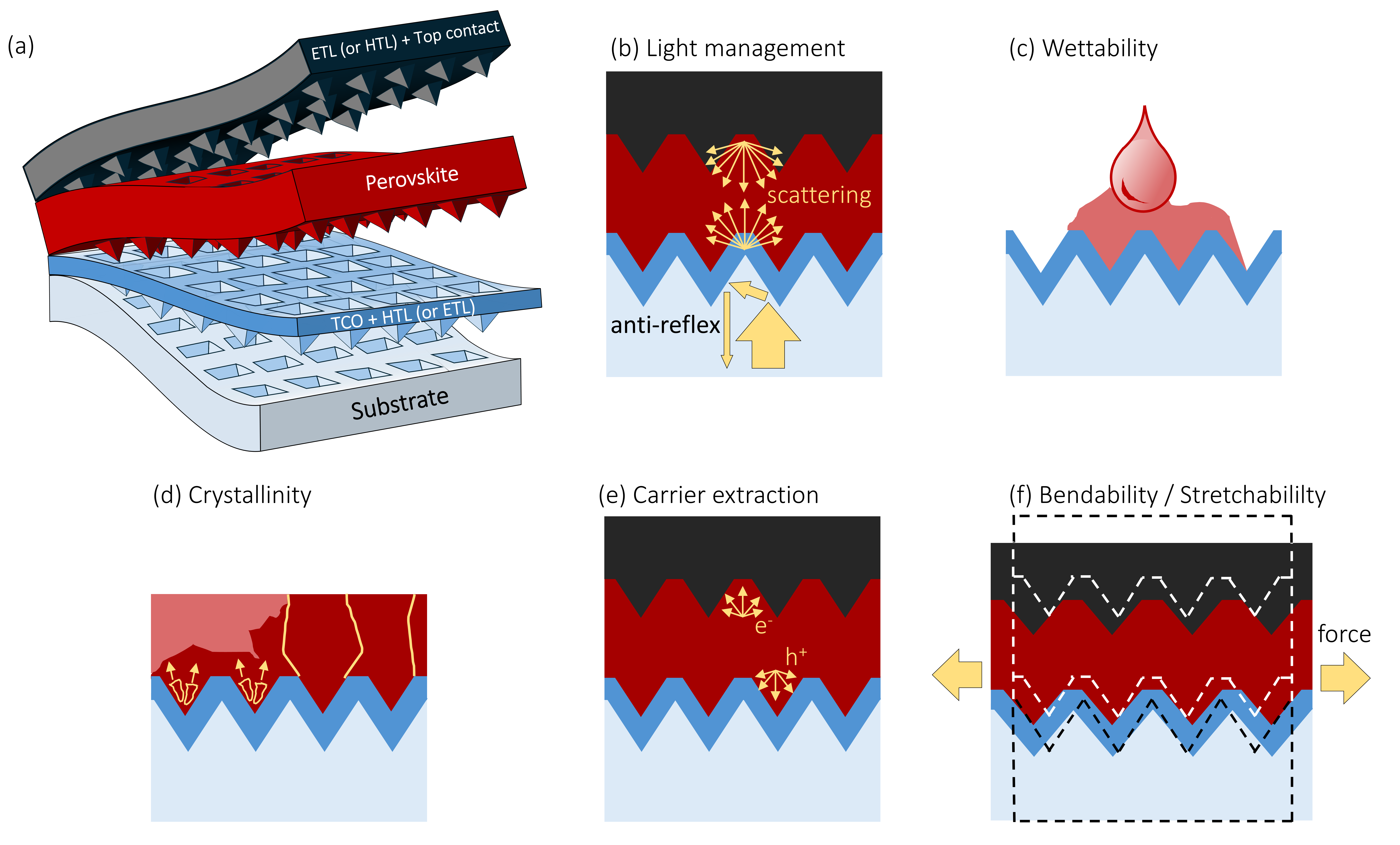}
  \caption{\textbf{Overview of the multiple potential advantages of nano- and micro-textures in perovskite solar cells.} (a) Schematic structure of a textured perovskite solar cell device. (b-f) Technological fields, where textures can be beneficial: (b) Light management, here illustrated for the case of illumination through a textured, transparent substrate. (c) Wettability of the perovskite solution during spin-coating, blade-coating or slot-die coating. (d) Crystallinity with large grains, grain boundaries predominantly oriented perpendicular to the layer, and suppressed phase segregation. (e) Carrier extraction and electronic performance, and (f) Reduction of residual stress after fabrication and enhanced bendability and stretchability of flexible devices.}
  \label{fig:overview}
\end{figure}

Nano- and micro-textures are indispensable in solar cells to enable the highest power conversion efficiencies due to their abilitiy to trap light and reduce reflection-losses in a broad spectral range. In the historic field of silicon thin-film solar cells, nano- or micro-textures were shown to strongly improve the optical performance on the one side. However, as collateral drawback, textures could at the same time induce lower quality bulk materials or interfaces, with macroscopic cracks in the layers, stress induced degradation, higher defect densities and uneven \textit{p-n}-junctions leading to more charge carrier recombination, which degraded the solar cell performance \cite{demant2015microcracks,cousins2006minimizing,zhong2013influence,matsui2002influence,mcintosh2009recombination, Soederstroem_2008,Python_2009}. Emerging perovskite solar cells have undergone rapid development in recent years, with power conversion efficiencies of around 27\% for single-junction devices and 35\% for tandem solar cells \cite{Green_2026}. Textures have played a major role in this technological development, partly in single junction (\textbf{Figure\,\ref{fig:overview}}a) but especially in tandem devices, as they are able to minimize broadband reflection losses and maximize light trapping (Figure\,\ref{fig:overview}b). Similarly as in silicon thin-film solar cell technologies, the implementation of textures for light management remained a challenging task with mixed blessings for a long time. In this case, the main concern for the implementation of textures in new generation perovskite solar cells was related to the complete coverage and compatible growth of the active layer, particularly if a solution processing method was employed. However, such limitation has been proven to be adequately overcome by choosing compatible processing conditions and texture characteristics \cite{hou2020efficient, tockhorn_nano-optical_2022, de2023monolithic, farag2023mitigation, winarto_periodic_2025}. In other studies, some characterization of textured perovskite suggested related issues that may compromise the optimal device performance such as inhomegeneous spatial perovskite composition, different charge carrier dynamics or detrimental stress on the perovskite layer \cite{lee2020perovskites,hou2020efficient,de2023monolithic}. One of the main tasks of scientists and technologists has been to address this trade-off and to find workarounds to minimize these collateral damages and implement textured perovskites layers that enhance the overall performance of the solar cell device. While working on this approach, more and more studies have recently reported on perovskite-based opto-electronic devices with the surprising property that, under certain circumstances, \textit{collateral advantages} are achieved upon implementation of nano- and micro-textures (Figure\,\ref{fig:overview}c-f). Nearly all of these studies giving account of textures in perovskite-based solar cells initially aimed at improving light management. However, some of these articles -- by the side -- also reported on additional effects. These effects go beyond the optics of the solar cell but can be as important as light harvesting to fabricate the optimal perovskite photovoltaic device. 
\\

To explore the possible \textit{collateral advantages} of textures in perovskite solar cells, here in this contribution we focus on the studies which addressed those benefits apart from light trapping, and classify them by technological fields. After reviewing the current bibliography, we found that such classification could be divided in (1) improved wetting of the textured surface by the perovskite solution, (2) enhanced perovskite crystallinity in terms of grain size,  preferential crystal orientation, and suppressed phase segregation, (3) improved charge carrier extraction and electronic performance of textured perovskite devices, and (4) increased mechanical robustness by reducing residual and external stress by texturing.


\section{Texture-related technological advantages beyond light management}

Concerning textures and photonic structures for light management, numerous review and perspective articles already exist in the fields of perovskite solar cells (e.g. \cite{Berry_2022,Jung2023,liu_light_2026}). Inspired by a progress report from the field of perovskite light emitting diodes, where textured perovskite morphology was not only discussed in the context of light management, but also with respect to carrier dynamics \cite{Zhong_2020}, here, we focus on recent articles on texture-related benefits beyond light management in perovskite solar cells.

\subsection{Textures for improved perovskite solution wetting}

\begin{figure}
  \includegraphics[width=\linewidth]{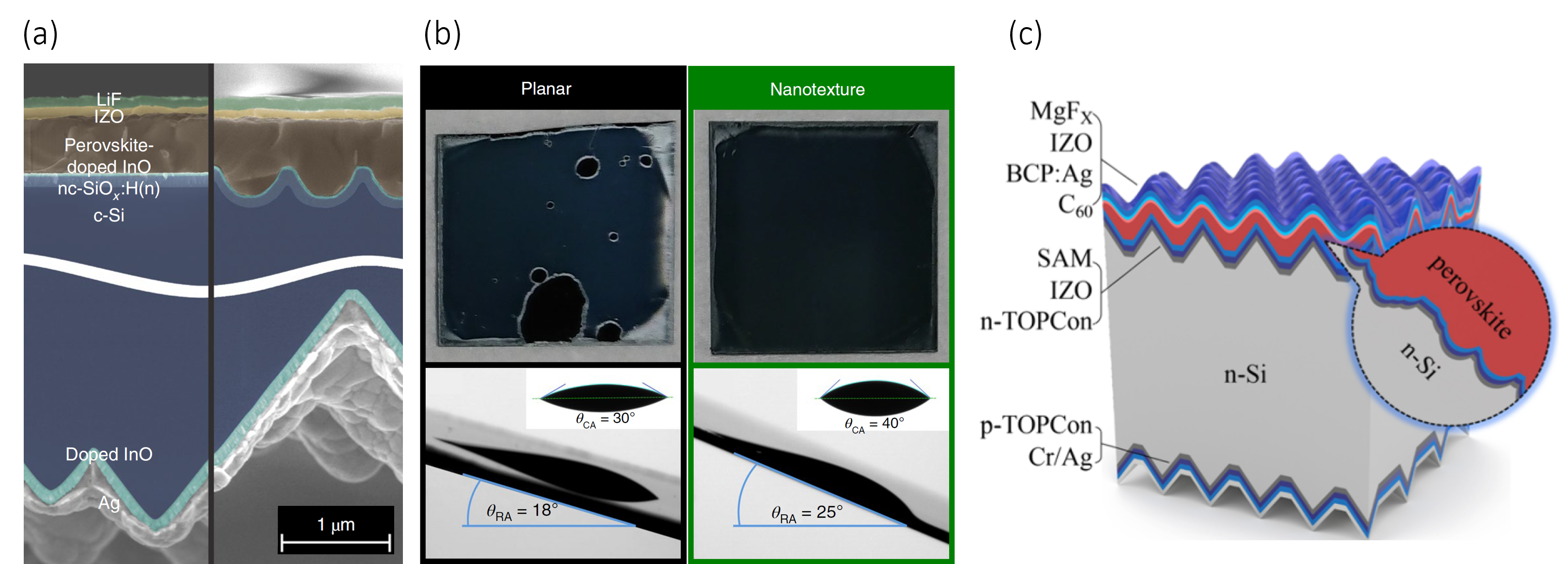}
  \caption{\textbf{Texture-enhanced perovskite solution wetting.} (a) Cross section scanning electron micrograph (SEM) of a perovskite-silicon tandem solar cell with sinusoidal nanotexture. (b) Spin-coated perovskite films on planar (left) and nanotextured (right) silicon bottom cells covered by Me-4PACz, respectively with side-view photographs of perovskite solution droplets indicating roll-off angle (RA) and static contact angle (CA; inset). (c) Schematic view of a hierarchical micro/nanostructured perovskite/silicon tandem solar cell. (a-b) Adapted from ref.\,\cite{tockhorn_nano-optical_2022}, 2022, Springer Nature, under the terms of the CC BY 4.0 license. (c) Reproduced from ref.\,\cite{ying_hierarchical_2024}, 2024, American Chemical Society, with permission under licence number 6231820280734.}
   \label{fig:wetting}
\end{figure}

One of the main challenges often associated with the use of textured substrates for the fabrication of perovskite solar cells is the complete and efficient coverage of the texture by the perovskite layer, especially when a solution processing method is employed. Wetting of solution-processed perovskite layers was mainly discussed in the context of perovskite-silicon tandem solar cells. A textured interface between perovskite top and silicon bottom cells is indispensable to reduce reflective light losses and to enhance light trapping in the silicon junction, thereby promoting the highest power conversion efficiencies. Beside the light trapping effect, Tockhorn \textit{et al.} \cite{tockhorn_nano-optical_2022} observed that sinusoidal nanotexturing of the silicon surface in the tandem device (\textbf{Figure\,\ref{fig:wetting}}a) greatly improved wetting of perovskite cell layers during spin-coating (Figure\,\ref{fig:wetting}b). A self-assembled monolayer (SAM) denoted Me-4PACz ([4-(3,6-dimethyl-9H-carbazol-9-yl)butyl]phosphonic acid) was used as hole contact layer, on which the perovskite film grew. Nanotexturing could reduce the share of unusable samples exhibiting macroscopic holes in the perovskite film from around 50\% for flat substrates to less than 5\% for nanotextured substrates, thus substantially enhancing the fabrication yield. Larger roll-off angles of the perovskite solution indicated an improved ability of the nanotextured surface to retain the perovskite solution compared to planar surfaces. Alternatively, Farag \textit{et al.} solved the issue of unsatisfactory perovskite wetting on planar Me-4PACz layers by fabricating the SAM via evaporation instead of spin-coating \cite{farag_evaporated_2023}.
A similar phenomenon was observed by Ying \textit{et al.} \cite{ying_monolithic_2022} from a black-silicon texture in monolithic perovskite-silicon tandems, which not only increased broadband light trapping, but also facilitated the wetting of perovskite. The same group showed later that a hierarchical micro-nanostructured pyramidal texture allows the deposition of fully textured perovskite absorbers growing conformally from a  solution-process in perovskite-silicon tandem solar cells \cite{ying_hierarchical_2024}. They explained the extraordinary wettability with the nanotextures acting as “blemishes” on the chemically homogeneous micropyramid surface (Figure\,\ref{fig:wetting}c). The concept of hierarchical textures was also taken up by Ayvazyan reporting on micrometer-sized pyramids with nano-sized black-silicon needles \cite{ayvazyan_hierarchical_2023}. Here, it was shown that a hierarchical surface on silicon substrates enables the formation of high-quality perovskite layers and contributes to a significant reduction in optical reflection losses. Turkay \textit{et al.} \cite{turkay_beyond_2025} showed that protrusion-free wetting of perovskite films from solution is also possible on around two-micrometer sized random pyramids by careful tailoring the film thickness to match the pyramids’ height profile, and by using a silicon oxide nanoparticle interlayer, which might play a similar role for wetting like the "blemishes" mentioned above. Winarto \textit{et al.} \cite{winarto_periodic_2025} reported on fully-textured conformal perovskite films fabricated via spin and slot-die coating on micrometer large periodic inverted pyramids, indicating that for specific texture geometries with consistent texture height, even a non-hierarchical "micro-texture-only" can lead to the required wetting surface properties to grow conformal perovskite layers.

\begin{figure}
\centering
  \includegraphics[width=0.9\linewidth]{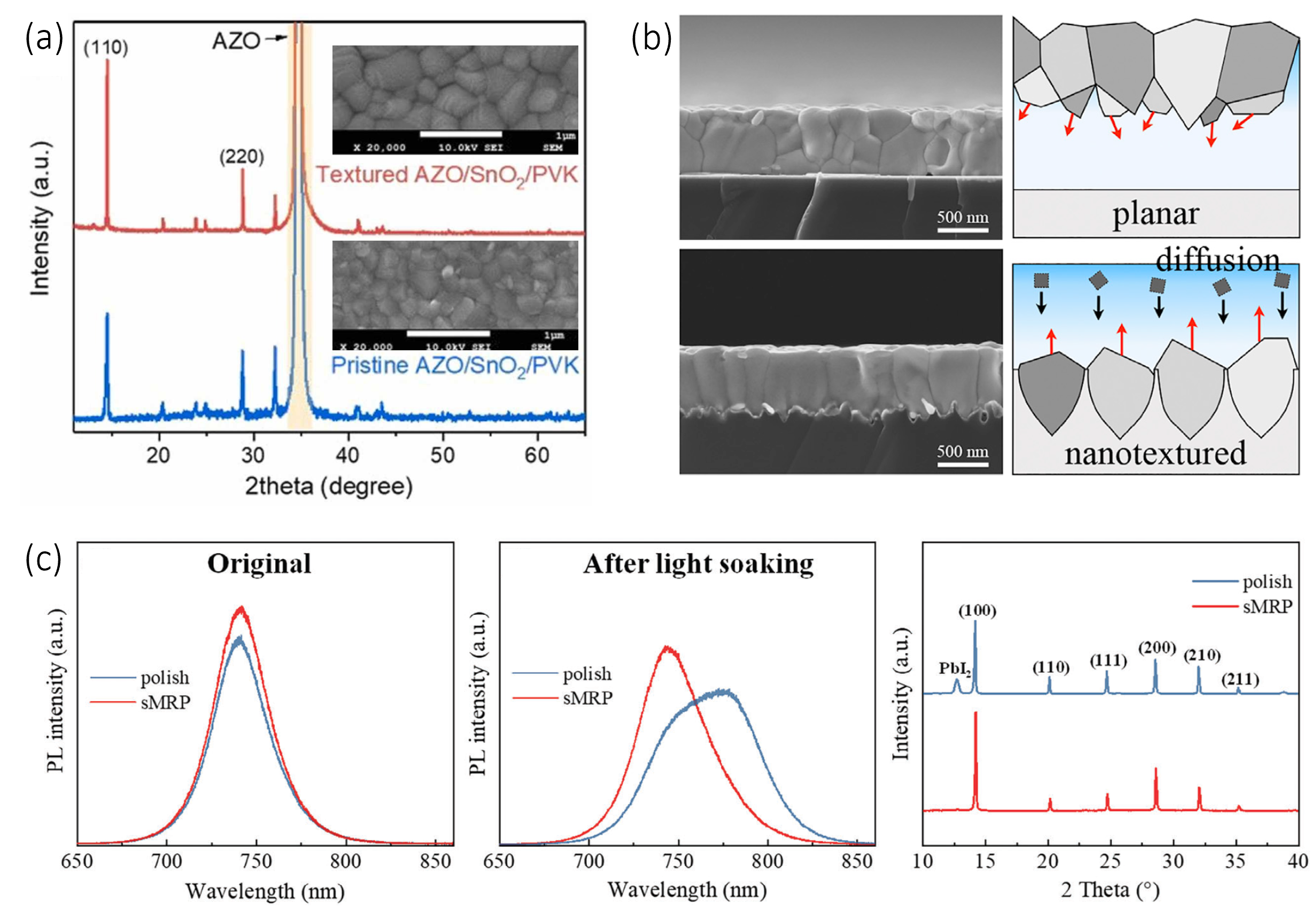}
  \caption{\textbf{Texture-enhanced perovskite crystal growth.} (a) X-ray diffraction spectra and corresponding scanning electron microscopic top view images (inset) of perovskite layers deposited on a textured (up) and a planar (down) aluminum doped zinc oxide (AZO) layer. (b) Cross-sectional SEM images (left) and schematic diagrams of the crystallization mechanism of perovskite films grown on planar (upper row) and  nanotextured (lower row) surfaces. (c) PL spectra before and after light soaking (left, middle) and XRD spectra (right) of perovskite layers deposited on polished (blue) and on sub-micrometer random pyramidal (sMRP) textured (red) substrates. (a) Adapted from ref.\,\cite{wang2023wet}, 2023, Elsevier, with permission under the licence number 6237030783676. (b) Reproduced from ref.\,\cite{ying_monolithic_2022}, 2022, Elsevier, with permission under licence number 6231830606087. (c) Adapted from ref.\,\cite{jiang_advancing_2024}, 2024, Wiley-VCH GmbH, with permission under the licence number 6237130153806.}
   \label{fig:crystalgrowth}
\end{figure}

\subsection{Textures for improving perovskite crystallinity}
Textures can also have a positive effect on perovskite crystallinity by promoting growth of larger perovskite grains (\textbf{Figure\,\ref{fig:crystalgrowth}}a), a preferential crystal orientation with vertical grain boundaries (Figure\,\ref{fig:crystalgrowth}b), and reduced lead halide segregation (Figure\,\ref{fig:crystalgrowth}c). An enlargement of perovskite crystal grains has been reported in the context of various textures, such as black-silicon textures \cite{ying_monolithic_2022}, sub-micrometer random pyramids \cite{jiang_advancing_2024}, inverted pyramidal micro-textures \cite{winarto_periodic_2025}, and textured aluminum doped zinc oxide layers \cite{wang2023wet}, when compared with their planar counterparts. In the latter study by Wang \textit{et al.}, it was observed that the morphology of the perovskite was considerably different when it was grown on a textured substrate originated from a textured aluminum zinc oxide electrode (Figure\,\ref{fig:crystalgrowth}a). The grains grew on average twice as large on the textured substrate compared to the flat surface\cite{wang2023wet}, which might be connected to the fact the the binding energy is directly dependent on the roughness of a sample \cite{dirksen1991fundamentals}. Narrower x-ray difrraction (XRD) peaks were also observed, which was interpreted as reduced scattering from grain boundaries. Furthermore, the study from Ying \textit{et al.} mentioned before showed how the textured surface acted as a nanoconfining scaffold to guide the vertical grain growth of the perovskite. This resulted in vertically aligned grain boundaries as illustrated in Figure\,\ref{fig:crystalgrowth}b, thus mitigating undesired carrier recombination losses \cite{ying_monolithic_2022}. Jiang \textit{et al.} reported on sub-micron random pyramidal (sMRP) textured surfaces facilitating the formation of a high-quality perovskite films with larger grain sizes and fewer internal pinholes compared to the polished flat counterpart. The photoluminescence (PL) spectra before and after 20 mins of blue light soaking showed a distortion on the planar perovskite emission upon illumination, while the textured one remained unchanged, leading to a reduced phase segregation and a more stable perovskite. Moreover, an XRD analysis also showed a preferred (100) crystal orientation and a removal of the PbI$_2$ signal enhancing the overall crystallinity of the perovskite layer \cite{jiang_advancing_2024}. These effects are shown in Figure\,\ref{fig:crystalgrowth}c.  As it was observed in other studies, \cite{winarto_periodic_2025, farag2023mitigation, yu_efficient_2026} the overall composition and crystallinity of the perovskite layer, which influence the optimal solar cell performance, have also been affected by the use of a textured substrate, impacting the final PbI$_2$ content and preferential crystal planes. Regarding the crystalline structure, some lattice parameters of the perovskite crystallites might be distorted when the layer is deposited on a textured surface, leading to a strain or stress degree \cite{liu2024strained, zhang2025coupled}, which sometimes can be introduced to positively influence the performance and stability of the devices \cite{jiang2024surface}.

\begin{figure}
\centering
  \includegraphics[width=0.9\linewidth]{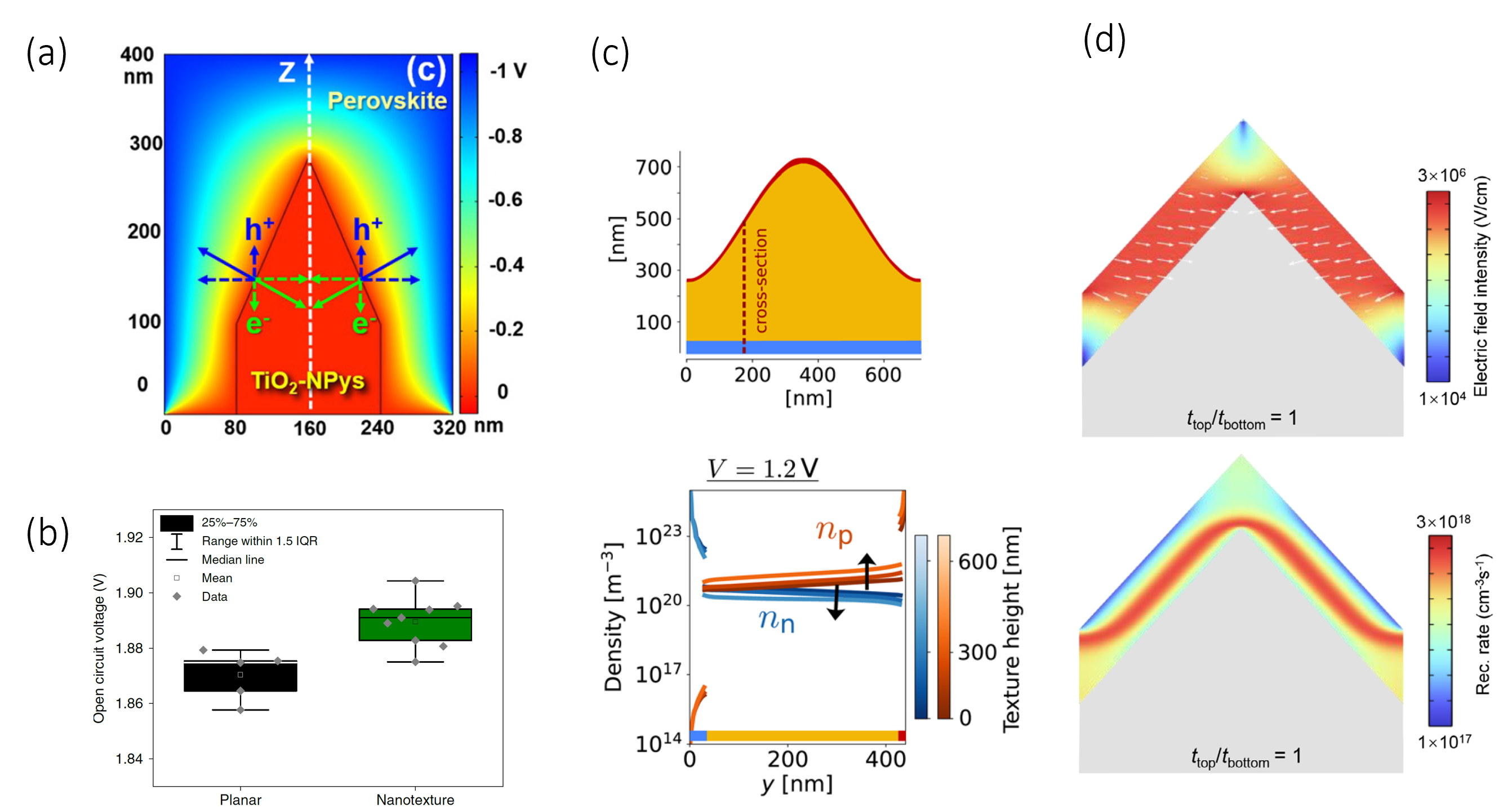}
  \caption{\textbf{Texture-enhanced electronic perovskite solar cell performance.} (a) Electric potential contours for titanium oxide nanopyramids in perovskite under reverse bias. (b) Box plot of the open-circuit voltage for planar and nanotextured perovskite-silicon tandem solar cells. (c) Geometry of the simulated 2D device with the vertical cross-section indicated (top), along which electron ($n_\mathrm{n}$) and hole ($n_\mathrm{p}$) densities (bottom) at an applied forward bias voltage close to the maximum power point for varying texture height. The texture-induced dysbalance of $n_\mathrm{n}$ and $n_\mathrm{p}$ might serve as possible explanation of the nanotexture-enhanced open-circuit voltage show in part (b). (d) Spatial electric field (top) carrier recombination rate (bottom) distributions within the perovskite top cell of a tandem device for 0.95 V forward bias voltage. (a) Reproduced from ref.\,\cite{lv_highefficiency_2020}, 2020, Wiley-VCH GmbH, with permission under the licence number 6231840319248. (b) Reproduced from ref.\,\cite{tockhorn_nano-optical_2022}, 2022, Springer Nature, under the terms of the CC BY 4.0 license. (c) Reproduced from ref.\,\cite{abdel_how_2026}, 2026, Royal Society of Chemistry, under the terms of the CC BY 3.0 license. (d) Reproduced and adapted from ref.\,\cite{yu_efficient_2026}, 2026, Wiley-VCH GmbH, with permission under the licence number 6231840825096.}
   \label{fig:electronics}
\end{figure}

\subsection{Textures for improved electronic performance}

Already in the early stages of perovskite solar cell development, it was observed that texturing of the titanium dioxide electrode led to an improved electron transfer in the device. Titanium dioxide architectures like nanocones \cite{zhong_synthesis_2015}, nanowires \cite{yu_development_2015}, and nanopyramids \cite{lv_highefficiency_2020} were discussed. \textbf{Figure\,\ref{fig:electronics}}a shows a nanopyramid array/perovskite system exhibiting an oriented electric field that can increase charge separation and accelerate charge transport, thereby suppressing charge recombination \cite{lv_highefficiency_2020}. Later, in several experimental studies on perovskite single-junction \cite{tockhorn_improved_2020} and perovskite-silicon tandem solar cells \cite{hou_efficient_2020,tockhorn_nano-optical_2022, zheng_balancing_2023} the implementation of textures is associated with a surprisingly large increase of open-circuit-voltage in the range of +15 mV to even +45 mV, i.e. larger than the expected logarithmic dependence on short-circuit current-density increase (Figure\,\ref{fig:electronics}b). In most of these studies this benefit remains unexplained. Hou \textit{et al.} proposed an explanation via a widening of the depletion region and suppressed non-radiative recombination in the texture valleys, resulting in improved charge carrier collection \cite{hou_efficient_2020}. The comprehensive numerical study of Abdel \textit{et al.} comes to a slightly different conclusion. Their results confirm the experimental results and explain the texture-related increased open-circuit voltage by a geometry effect. They show that texturing one of the absorber/transport layer interfaces increases the imbalance between electron and hole densities in the absorber (Figure\,\ref{fig:electronics}c, bottom), thereby reducing non-radiative recombination, and hence enabling larger open-circuit voltages \cite{abdel_how_2026}. Yu \textit{et al.} observe that a conformal perovskite morphology can partially suppress carrier recombination, especially at the pyramid valleys (Figure\,\ref{fig:electronics}d, \cite{yu_efficient_2026}). Time resolved PL measurements of perovskite layers deposited on different microtextured random pyramids showed a longer lifetime of charge carriers as the texture becomes more pronounced, which might be an indicator of less recombination probability within the perovskite layer \cite{de2023monolithic}. Unfortunately, in this case, the overall tandem open-circuit voltage was not shown to be consistently enhanced, which might be due to other recombination losses in the multijunction structure promoted by rougher textures.

\begin{figure}
\centering
  \includegraphics[width=\linewidth]{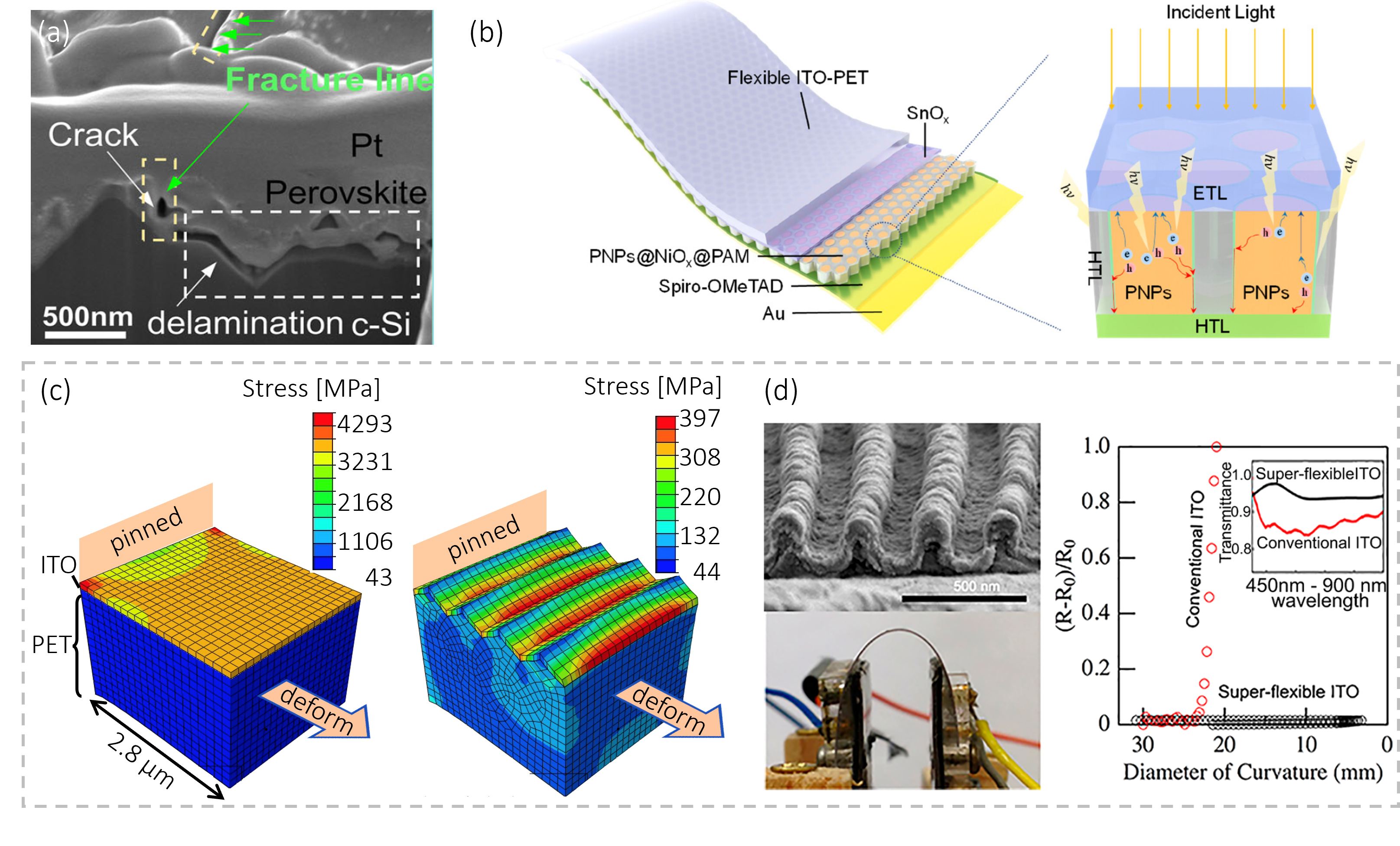}
  \caption{\textbf{Impact of nano- and micro-textures on mechanical stress.} (a) Cross section scanning electron micrograph of a flexible perovskite silicon tandem solar cell illustrating the problems that can occur upon bending (cracking, delamination). (b) Structure diagram and cross-sectional schematic diagram of a flexible perovskite nanopillar-based solar cell.  Results shown in (c-d), surrounded by a gray dashed line, are not directly related to perovskite solar cells, but support the use of textures as possible stress releasing strategy: (c) Simulated von Mises stress of thin ITO layers on flat (left) and micro-textured (right) PET substrates with deformation indicated by orange arrows. (d) ITO periodic grating patterns on the surfaces of the polymer substrates (top), illustration of the bending configuration (bottom) and, and relative resistance increase as a function of bending curvature. (a) Reproduced from ref.\,\cite{sun_flexible_2025}, 2025, Springer Nature, under the terms of the CC BY 4.0 license. (b) Reproduced from ref.\,\cite{zhu2022three}, 2022, American Chemical Society, with permission under the licence number 6281980749044. (c) Reproduced and adapted from ref.\,\cite{rieckhoff_texture-enhanced_2025}, 2025, Wiley-VCH GmbH, under the terms of the CC BY 4.0 license. (d) Reproduced from ref.\,\cite{dong_superflexibility_2018}, 2018, Wiley-VCH GmbH, with permission under the licence number 6231960284188.}
   \label{fig:mechanics}
\end{figure}

\subsection{Textures for reduced mechanical stress}

Residual stress in the perovskite layers after processing is often a performance limiting factor of the final solar cell device. The problem is further exacerbated when external mechanical forces are applied, for instance in case of flexible opto-electronic devices experiencing bending or stretching. \textbf{Figure\,\ref{fig:mechanics}a} shows the example of a flexible perovskite-silicon tandem solar cell after bending, exhibiting cracks and delamination of the perovskite film. In this and in other studies the detrimental effect of mechanical stress was successfully mitigated by careful adaption of the perovskite composition \cite{sun_flexible_2025,yu_efficient_2026}, or by optimization of contact and recombination layers \cite{wang_flexible_2026}. Zhu \textit{et al.} presented a different stress-mitigating approach for flexible perovskite solar cells by using a flexible substrate with 500\,nm periodic perovskite nanopillars grown on a SnO$_2$ flat interface (Figure\,\ref{fig:mechanics}b). Here, the textured perovskite solar cells exhibited a strongly enhanced mechanical robustness upon bending compared to a conventional planar control device, supported by stress-strain simulations \cite{zhu2022three}. Apart from this notable example, the impact of texturing on mechanical stress in perovskite solar cells was only marginally discussed, so far. 

In contrast, in related fields of opto-electronics, we found various articles discussing how texturing can positively influence the mechanical robustness. Wang \textit{et al.} observed a double effect in perovskite photodetectors on textured plastic foils, with hierarchical micro-nanostructures at the rear side of the foil substrate, enhancing the optical performance and improving the mechanical robustness upon bending \cite{wang_back-reflected_2020}. Ginting \textit{et al.} developed flexible inverse micro-cone arrays for organic and perovskite solar cells strongly increasing light trapping. They also found indications that the textured organic solar cells exhibited a better mechanical robustness upon bending compared to their flat counterparts \cite{ginting_dual_2018}. A similar effect was reported by Lin \textit{et al.} using nanocones in flexible amorphous silicon solar cells  \cite{lin2016high}. The examples shown in Figure\,\ref{fig:mechanics}(c-d) deal with the configuration of a brittle thin film with high Young's modulus deposited on a soft substrate with much lower Young's modulus -- a situation that usually occurs in flexible perovskite solar cells at the interface between the flexible substrate (e.g PET) and the transparent electrode, normally inconveniently brittle, like indium tin oxide (ITO). Rieckhoff \textit{et al.} demonstrated that the bending durability of ITO transparent electrodes on plastic foils can massively increase if the foils exhibited a grating micro-texture. Simulations attributed this texture-induced robustness upon bending to strongly reduced stress in the ITO layer (Figure\,\ref{fig:mechanics}c). Given the fabrication method, the grating surface presented a nano-wrinkled surface composing a hierarchical micro-nano-texture that may have further contributed to a better mechanical performance \cite{rieckhoff_texture-enhanced_2025}.  A resembling study by Dong \textit{et al.} \cite{dong_superflexibility_2018} using also a grating-like structure came to a similar conclusion in which this texture was employed to retain the conductivity and optical qualities of the ITO (Figure\,\ref{fig:mechanics}d). Using a different texturing approach that consisted of implementing 2D meshed ITO configurations, Sakamoto \textit{et al.} \cite{sakamoto_highly_2018} showed that such textures contributed to a lower tensile stress and hindered crack propagation during bending.

\begin{table}
\caption{Summary of studies reporting on textures in perovskite solar cells loosely sorted by texture-related collateral technological advantages in addition to light management. The light management aspect plays a role, often the primary role, in all the mentioned studies, but it is not explicitly mentioned here in this table.}
\small
    \centering
    \begin{tabular}[t]{clcccccl}
    \hline
     & & & \multicolumn{4}{c}{\thead{Texture-related \textit{collateral advantage} \\in addition to light management regarding...}}&  \\
     & \thead{Author / year} & \thead{ref.} & \thead{wetting} & \thead{crystallinity} & \thead{electronics} & \thead{mechanics} & \thead{Texture details}  \\
    \hline
     \parbox[0pt][2.0em][c]{0cm}{} \rownumber & Tockhorn 2022 & \cite{tockhorn_nano-optical_2022} & x & & x & & sub-micrometer sinusoidal textures \\
     \parbox[0pt][2.0em][c]{0cm}{} \rownumber & Ying 2022 & \cite{ying_monolithic_2022} & x & x & & & black-silicon texture \\
     \parbox[0pt][2.0em][c]{0cm}{} \rownumber & Ying 2024 & \cite{ying_hierarchical_2024} & x & & & & \makecell[l]{hierarchical micro-nano-textures: \\micro-pyramids with nano-"blemishes"} \\
     \parbox[0pt][2.0em][c]{0cm}{} \rownumber & Ayvazyan 2023 & \cite{ayvazyan_hierarchical_2023} & x & & & & \makecell[l]{hierarchical micro-nano-textures: \\micro-pyramids with black-Si texture} \\
     \parbox[0pt][2.0em][c]{0cm}{} \rownumber & Turkay 2025 & \cite{turkay_beyond_2025} & x & & & & \makecell[l]{hierarchical nano-micro-textures: \\ micro-pyramids with nanoparticle coverage} \\
     \hline
     \parbox[0pt][2.0em][c]{0cm}{} \rownumber & Wang 2023 & \cite{wang2023wet} & x & x & & & textured aluminum doped zinc oxide \\
     \parbox[0pt][2.0em][c]{0cm}{} \rownumber & Jiang 2024 & \cite{jiang_advancing_2024} & x & x & & & sub-micrometer pyramids \\
     \parbox[0pt][2.0em][c]{0cm}{} \rownumber & Winarto 2025 & \cite{winarto_periodic_2025} &  & x & & & periodic, inverted micro-pyramids \\
     \parbox[0pt][2.0em][c]{0cm}{} \rownumber & Yu 2026 & \cite{yu_efficient_2026} & (x) & x & x & (x) & random micro-pyramids \\
     \hline
     \parbox[0pt][2.0em][c]{0cm}{} \rownumber & Lv 2020 & \cite{lv_highefficiency_2020} & & & x & & nanopyramid titanium dioxide electrode \\
     \parbox[0pt][2.0em][c]{0cm}{} \rownumber & Tockhorn 2020 & \cite{tockhorn_improved_2020} & & & x & & sub-micrometer sinusoidal textures \\
     \parbox[0pt][2.0em][c]{0cm}{} \rownumber & Hou 2020 & \cite{hou_efficient_2020} & & & x & & sub-micrometer random pyramids \\
     \parbox[0pt][2.0em][c]{0cm}{} \rownumber & Zheng 2023 & \cite{zheng_balancing_2023} & & & x & (x) & random micro-pyramids \\
     \parbox[0pt][2.0em][c]{0cm}{} \rownumber & Abdel 2026 & \cite{abdel_how_2026} & & & x & & \makecell[l]{sub-micrometer sinusoidal textures \\(numerical study)} \\
     \hline
     \parbox[0pt][2.0em][c]{0cm}{} \rownumber & Zhu 2022 & \cite{zhu2022three} & & &  & x & \makecell[l]{sub-micrometer periodic nanopillars} \\
     \parbox[0pt][2.0em][c]{0cm}{} \rownumber & Sun 2025 & \cite{sun_flexible_2025} & & &  & (x) & \makecell[l]{random micro-pyramids} \\
     \parbox[0pt][2.0em][c]{0cm}{} \rownumber & Wang 2026 & \cite{wang_flexible_2026} & & &  & (x) & \makecell[l]{sub-micrometer random pyramids} \\
     \parbox[0pt][2.0em][c]{0cm}{} \rownumber & \textcolor{gray}{Ginting 2018} & \textcolor{gray}{\cite{ginting_dual_2018}} & & &  & \textcolor{gray}{x} & \textcolor{gray}{\makecell[l]{inverse micro-cone arrays \\(organic solar cells)}} \\
     \parbox[0pt][2.0em][c]{0cm}{} \rownumber & \textcolor{gray}{Dong 2018} & \textcolor{gray}{\cite{dong_superflexibility_2018}} & & &  & \textcolor{gray}{x} & \textcolor{gray}{\makecell[l]{sub-micrometer grating \\ (transparent electrodes)}} \\
     \parbox[0pt][2.0em][c]{0cm}{} \rownumber & \textcolor{gray}{Sakamoto 2018} & \textcolor{gray}{\cite{sakamoto_highly_2018}} & & &  & \textcolor{gray}{x} & \textcolor{gray}{\makecell[l]{2D micro-pattern \\ (transparent electrodes)}} \\
     \parbox[0pt][2.0em][c]{0cm}{} \rownumber & \textcolor{gray}{Wang 2020} & \textcolor{gray}{\cite{wang_back-reflected_2020}} & & &  & \textcolor{gray}{x} & \textcolor{gray}{\makecell[l]{hierarchical micro-nano-textures \\(perovskite photodetectors)}} \\
     \parbox[0pt][2.0em][c]{0cm}{} \rownumber & \textcolor{gray}{Rieckhoff 2025} & \textcolor{gray}{\cite{rieckhoff_texture-enhanced_2025}} & & &  & \textcolor{gray}{x} & \textcolor{gray}{\makecell[l]{hierarchical micro-nano-grating \\(transparent electrodes)}} \\
    \hline
    \multicolumn{8}{l}{(x): Aspect is discussed in the paper, but not in context with textures.}\\
    \multicolumn{8}{l}{\textcolor{gray}{Gray: Texture-related mechanical enhancement reported in application field different from perovskite solar cells (in brackets).}}\\
    \hline
     
    \end{tabular}
    \label{tab:summary}
\end{table}

\subsection{Discussion}
A representative selection of studies discussing texture-related technological benefits in perovskite solar cells, which are different from light management, is summarized in \textbf{Table\,\ref{tab:summary}}. In all of these studies, the primary motivation to implement textures was improving the optical solar cell performance in terms of reflection-reduction and enhanced light harvesting. But all of them report, more or less by the side, on other phenomena that could enhance another aspect of the solar cell performance. Perovskite-solution wetting was shown to improve on nano-textured surfaces (rows 1-7 of Table\,\ref{tab:summary}). Particularly, so called hierarchical textures were discussed to be very promising. They are composed of a larger micro-texture optimized for broadband light management, and a superimposed nano-texture that enables conformal wetting of the perovskite precursor solution during deposition (rows 3-5 of Table\,\ref{tab:summary}). Enhanced perovskite crystallinity in terms of grain size, crystal orientation and phase homogeneity was reported for various different textures ranging from nano- to micro-scale (rows 6-9 of Table\,\ref{tab:summary}). Also for enhanced electronic (rows 10-14 of Table\,\ref{tab:summary}) and mechanical (rows 15-22 of Table\,\ref{tab:summary}) performance still no trend was found to emerge concerning optimum texture-dimensions. 
The reported additional texture-related phenomena in perovskite solar cells are often not independent of each others but are often closely correlated: For instance, textures can not only promote wetting of the perovskite precursor solution during the coating process, but also initiate crystallization starting from the substrate enabling larger grains and vertical grain boundaries \cite{ying_monolithic_2022}. Textures can diminish phase segregation in the perovskite films, hence reducing the observed amount of PbI$_2$ \cite{jiang_advancing_2024,winarto_periodic_2025}. Perovskite composition and phase homogeneity are in turn found to be closely related to residual stress in the film \cite{sun_flexible_2025,yu_efficient_2026}. 
\\
While textures are meanwhile used by default in perovskite-silicon tandem solar cells, they are still less frequently reported in perovskite single-junction and all-perovskite tandem solar cells. Due to the multi-physical and striking advantages of nano- and micro-textures summarized in this article, we believe that future high-performance perovskite-based opto-electronic devices - including solar cells, photodetectors and light emitting diodes - will comprise textures. Proper design of textures in perovskite opto-electronic devices will become a multi-physical, multi-parameter optimization task in future work, where optical, electrical and mechanical aspects must be incorporated. Particularly in the field of textured, flexible perovskite solar cells \cite{soldera_toward_2020, blackburn_back-contact_2025, sun_flexible_2025, wang_flexible_2026} we see a great performance and stability enhancing potential via optimized nano- and micro-texturing.

\section{Conclusion}

We identified a potential new trend in perovskite solar cells to simultaneously enhance multiple aspects of device performance by implementation of nano- and micro-textures. We reviewed studies from the very recent past reporting on perovskite solar cells with the surprising property that collateral technological advantages are achieved upon nano- and micro-texturing, going far beyond the well-known light management effect. We classified the studies according to their texture-related additional advantages, which are (1) improved wetting of perovskite films from solution on textured surfaces, (2) enhanced crystallinity in terms of grain size, crystal orientation, and phase homogeneity, (3) improved charge carrier extraction and electronic performance, and (4) increased mechanical robustness by reducing residual and external stress. The findings are of potential relevance not only for perovskite solar cells, but generally for opto-electronic thin-film devices based on other materials such as organic or inorganic semiconductors. We believe that nano- and micro-textures will become an integral part of future opto-electronic devices with superior optical, electrical and mechanical performance, and see the highest potential for flexible applications.


\medskip

\medskip
\textbf{Acknowledgements} \par 
The authors thank for funding from the Helmholtz Association within the HySPRINT Innovation lab project, and the project “Zeitenwende–Zukunftstechnologie Tandem–Solarzellen”. The authors further thank for funding from the European Innovation Council (EIC) within the European project JUMP-INTO-SPACE (grant agreement No 101162377).

\medskip

%
\bibliographystyle{MSP}
\bibliography{Bibliography_4-Mar-2026}

\newpage


\begin{figure}
\textbf{Table of Contents}\\
\medskip
  \includegraphics[width=55 mm]{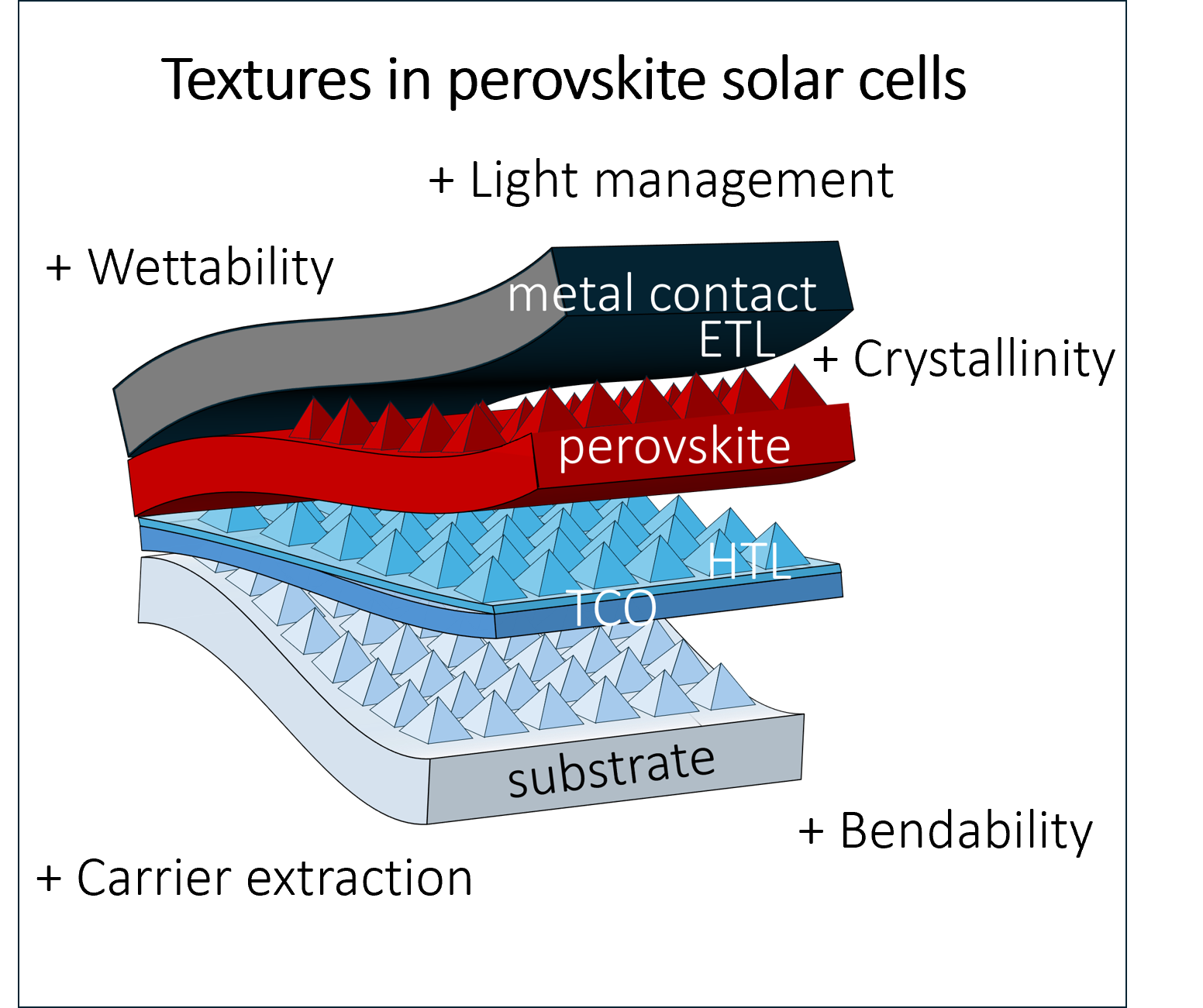}
  \medskip
  \caption*{Nano- and micro-texture can improve the performance of perovskite solar cells in various ways. In addition to the well-known effect of improved light management, nano- and micro-textures have demonstrated additional benefits in the fields of wettability, crystallinity, carrier extraction, and mechanical bendability. This paper summarizes recently published studies and discusses the promising prospects for perovskite-based, textured opto-electronic thin-film devices.}
\end{figure}

\end{document}